 \documentclass[final,5p,times,twocolumn]{elsarticle}

\usepackage{amssymb}
\usepackage{lipsum}
\usepackage{hyperref}
\hypersetup{colorlinks,citecolor=blue,filecolor=blue,linkcolor=blue,urlcolor=blue}
\usepackage{graphicx}
\usepackage{dcolumn}
\usepackage{amsmath}
\usepackage{multirow}
\usepackage{threeparttable}
\usepackage{epstopdf}

\journal{Physics Letters B}

\begin{document}
\begin{frontmatter}



\title{Evolution of the nuclear spin-orbit splitting explored via  \\the $^{32}$Si($d$,$p$)$^{33}$Si reaction using SOLARIS}

\author[sustech,anl]{J.~Chen}
\ead{chenjie@sustech.edu.cn}

\author[anl]{B.~P.~Kay}
\author[anl]{C.~R.~Hoffman}
\author[anl]{T.~L.~Tang}
\author[anl]{I.~A.~Tolstukhin}
\author[msu]{D.~Bazin}
\author[msu]{R.~S.~Lubna}
\author[Spain]{Y.~Ayyad}
\author[msu]{S.~Beceiro-Novo}
\author[anu]{B. J. Coombes}
\author[Manchester,cern]{S.~J.~Freeman}
\author[liverpool]{L.~P.~Gaffney}
\author[msu]{R.~Garg}
\author[anl]{H.~Jayatissa}
\author[Davidson]{A.~N.~Kuchera}
\author[Manchester]{P.~MacGregor}
\author[anu]{A.~J.~Mitchell}
\author[msu]{W.~Mittig}
\author[msu,hope]{B.~Monteagudo}
\author[Spain]{A.~Munoz-Ramos}
\author[anl]{C.~Müller-Gatermann}
\author[infn,padova]{F.~Recchia}
\author[msu]{N.~Rijal}	
\author[msu]{C.~Santamaria}
\author[msu]{M.~Z.~Serikow}
\author[Manchester]{D.~K.~Sharp}
\author[Connecticut]{J.~Smith}
\author[Connecticut]{J.~K.~Stecenko}
\author[lsu]{G.~L.~Wilson}
\author[Connecticut]{A.~H.~Wuosmaa}
\author[Zhuhai]{C.~X.~Yuan}
\author[msu]{J.~C.~Zamora}	
\author[Zhuhai]{Y.~N.~Zhang }

\affiliation[sustech]{organization={College of Science, Southern University of Science and Technology},
           city={Shenzhen},
           postcode={518055}, 
           state={Guangdong},
           country={China}}
\affiliation[anl]{organization={Physics Division, Argonne National Laboratory},
           city={Lemont},
           postcode={60439}, 
           state={Illinois},
           country={USA}}
\affiliation[msu]{organization={Facility for Rare Isotope Beams, Michigan State University},
           city={East Lansing},
           postcode={48824}, 
           state={Michigan},
           country={USA}}
\affiliation[Spain]{organization={IGFAE, Universidade de Santiago de Compostela},
           city={Santiago de Compostela},
           postcode={E-15782}, 
           country={Spain}}
\affiliation[anu]{organization={Department of Nuclear Physics and Accelerator Applications, Research School of Physics, Australian National University},
           city={Canberra},
           postcode={ACT 2601}, 
           country={Australia}}
\affiliation[Manchester]{organization={Department of Physics, University of Manchester},
           city={Manchester},
           postcode={M13 9PL}, 
           country={United Kingdom}}
\affiliation[cern]{organization={EP Department, CERN},
           city={Geneva},
           postcode={CH-1211}, 
           country={Switzerland}}
\affiliation[liverpool]{organization={Department of Physics, University of Liverpool},
           city={Liverpool},
           postcode={L69 3BX}, 
           country={United Kingdom}}
\affiliation[Davidson]{organization={Department of Physics, Davidson College},
           city={Davidson},
           postcode={28035}, 
           state={North Carolina},
           country={USA}}
\affiliation[hope]{organization={Department of Physics, Hope College},
           city={Holland},
           postcode={49422-9000}, 
           state={Michigan },
           country={USA}}
\affiliation[infn]{organization={INFN, Laboratori Nazionali di Legnaro},
           city={Legnaro},
           postcode={I-35020}, 
           country={Italy}}
\affiliation[padova]{organization={Dipartimento di Fisica e Astronomia dell'Universitá di Padova},
           city={Padova},
           postcode={I-35131}, 
           country={Italy}}
\affiliation[Connecticut]{organization={Department of Physics, University of Connecticut},
           city={Storrs},
           postcode={06269}, 
           state={Connecticut},
           country={USA}}
\affiliation[lsu]{organization={Department of Physics and Astronomy, Louisiana State University}, 
           city={Baton Rouge},
           postcode={70803}, 
           state={Louisiana},
           country={USA}}
\affiliation[Zhuhai]{organization={Sino-French Institute of Nuclear Engineering and Technology, Sun Yat-Sen University},
           city={Zhuhai},
           postcode={519082}, 
           state={Guangdong},
           country={China}}

\begin{abstract}
The spin-orbit splitting between neutron 1$p$ orbitals at $^{33}$Si has been deduced using the single-neutron-adding ($d$,$p$) reaction in inverse kinematics with a beam of $^{32}$Si, a long-lived radioisotope. Reaction products were analyzed by the newly implemented SOLARIS spectrometer at the reaccelerated-beam facility at the National Superconducting Cyclotron Laboratory. The measurements show reasonable agreement with shell-model calculations that incorporate modern cross-shell interactions, but they contradict the prediction of proton density depletion based on relativistic mean-field theory. The evolution of the neutron 1$p$-shell orbitals is systematically studied using the present and existing data in the isotonic chains of $N=17$, 19, and 21. In each case, a smooth decrease in the separation of the $1p_{3/2}$-$1p_{1/2}$ orbitals is seen as the respective $p$-orbitals approach zero binding, suggesting that the finite nuclear potential strongly influences the evolution of nuclear structure in this region.

\end{abstract}




\end{frontmatter}




\section{Introduction}
\label{introduction}

The spin-orbit (SO) potential, which arises from the coupling of a particle’s orbital motion to its intrinsic spin, plays an important role in atomic~\cite{Soumyanarayanan} and nuclear~\cite{Mayer} structure. Incorporating a SO term in the nuclear potential is necessary to describe experimental data, which revealed enhanced stability at particular ``magic" nucleon numbers. The SO term lifts the degeneracy of orbitals with total nucleon angular momentum $j$ and creates a splitting between orbitals with $j=\ell+s$ and  $j=\ell-s$, where $l$ and $s$ are the orbital and spin angular momenta. 

Recently, the evolution of the energy separation between the neutron $1p_{3/2}$ and $1p_{1/2}$ SO partners along the $N=21$ isotones has received much attention~\cite{Mutschler,Burgunder,Kay,SorlinPLB,Grasso,Otsuka_review}. A sudden reduction in the separation of the neutron 1$p$ SO partners was suggested to occur between $^{37}$S and $^{35}$Si, speculated to be the consequence of a proton ``bubble'' in $^{34}$Si where the $1s_{1/2}$ proton orbital was measured to be almost empty relative to $^{36}$S where the $1s_{1/2}$ orbital is fully occupied. It is thus postulated that there is a central density depletion which results in a weakening of the two-body SO potential~\cite{Mutschler,Burgunder}. However, in these initial studies, only the 3/2$^-$ and 1/2$^-$ states, representing the dominant fragment of the $1p_{3/2}$ and $1p_{1/2}$ single-particle strengths, respectively, were used~\cite{Burgunder}. In contrast, a smooth reduction of the SO splitting in these $N=21$ isotones is obtained when taking into account the fragmentation of these single-neutron strengths. The smooth reduction was discussed in terms of the weak binding of these low-$\ell$ states, where the corresponding orbitals show ``lingering'' effect approaching the neutron-emission threshold~\cite{Kay}. However, this highly debated interpretation still calls for investigation on the question of whether the weak-binding effect or weakening of the two-body SO potential drives the change in the $1p$ SO-splitting change in this region~\cite{SorlinPLB,Grasso,Otsuka_review}.

Theoretically, the existence of a proton bubble structure within $^{34}$Si and its possible impact on the SO splitting is not yet well established. {\it Ab-initio} predictions regarding the existence of the bubble structure have been shown to vary significantly with the choice of Hamiltonian used~\cite{Duguet}. Relativistic mean-field (RMF) calculations suggest that the SO splitting weakens with enhanced pairing correlations, vibrational couplings, and model parameters~\cite{Yao, Karakatsanis}. 

In order to investigate whether a common mechanism is driving the evolution of the SO splitting in this region, we present new data and a systematic study of the $1p$ orbital single-particle energies (SPEs) for even $Z$ odd $N=17$-$21$ nuclei. In particular, knowledge of the change in SO splitting from S ($Z=16$) to Si ($Z=14$) is crucial to determine whether there is a sudden reduction of SO splitting due to the removal of the $1s_{1/2}$ protons in the core. In the present work, the SPEs of the neutron $1p$ and $0f_{7/2}$ orbitals in $^{33}$Si have been determined. In particular, the $1/2^-$ state carrying the dominant fragment of the $1p_{1/2}$ single-particle strength, which determines the SO splitting, has been observed for the first time.

In this study, a strikingly smooth evolution in the SO splitting is seen as the nuclei become less bound. Importantly, there is no significant deviation from this trend across any of the nuclei. This trend is reproduced by the Woods-Saxon calculations, which include data approaching zero neutron binding energy, indicating that the finite nuclear potential strongly influences the evolution of nuclear structure in this region. 

\section{Experiment}
\label{Experiment}

\begin{figure}
\includegraphics[width=2.0\columnwidth]{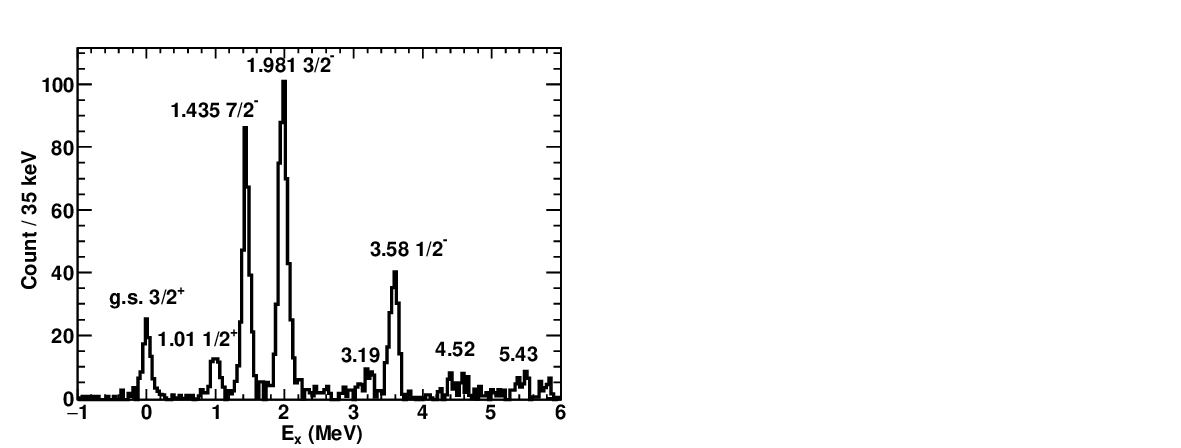}
\caption{\label{fig2} The excitation-energy spectrum for states in $^{33}$Si populated via the $^{32}$Si$(d,p)$ reaction. The peaks are labeled by their excitation energies (in MeV) together with their spin-parity assignments.}
\end{figure} 

The experiment to study $^{33}$Si was carried out at the ReA6 reaccelerator-beam facility of the National Superconducting Cyclotron Laboratory (NSCL). The 8.3-MeV/u $^{32}$Si beam, a long-lived radioisotope, had an intensity of approximately $10^5$ particles per second and a beam purity of $\sim 90\%$ due to a 32S contaminant. 
Protons produced by reactions of the $^{32}$Si beam impinging on a 120-$\mu$g/cm$^2$ (CD$_2)_n$ target were analyzed by the newly developed SOLARIS solenoidal spectrometer~\cite{SOLARIS} with a magnetic field of 3~T. SOLARIS is based on the solenoidal spectrometer concept pioneered at Argonne National Laboratory~\cite{Schiffer,Wuosmaa,Lighthall}, which was set up in a similar way to that described in Ref.~\cite{CHEN}. The energies and positions at which the protons returned to the beam axis were measured using the HELIOS four-sided array of position-sensitive silicon detectors (PSD). The projectile-like Si recoils were uniquely identified from other reaction channels or un-reacted beam components, including the $^{32}$S contaminant, by a square $(5 \times 5$ cm$^2$) Si recoil detector telescope with quadrant segmentation in the $\Delta E$. The recoil detectors were 53-$\mu$m and 150-$\mu$m thick, serving as $\Delta E$ and $E$ detectors, and were shielded from the primary beam by an 8-mm diameter blocker. 
A 20-ns timing coincidence between the protons and the Si recoils was applied to select the ($d$,$p$) reaction channel, to separate the S contamination and to reduce the background. 

\begin{figure}
\includegraphics[width=1.0\columnwidth]{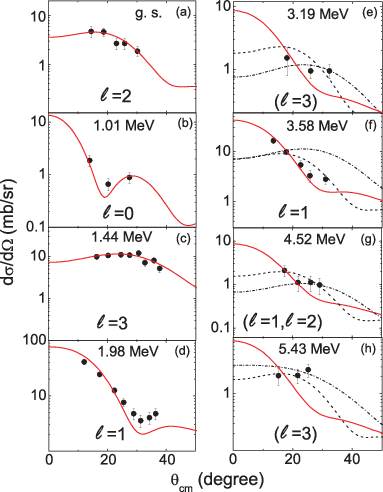}
\caption{\label{fig3} Differential cross sections from the $^{32}$Si$(d,p)^{33}$Si reaction for the low-lying states of $^{33}$Si. DWBA calculations are plotted as solid red lines with the $\ell$ values labeled for the known states (a-d). For the newly observed states (e-h), the $\ell=1$ (red solid lines), $\ell=2$ (black dashed lines) and $\ell=3$ (black dot-dashed lines) are plotted. The adopted $\ell$ values are labeled for each state.}
\end{figure}

Figure~\ref{fig2} shows the excitation-energy spectrum of $^{33}$Si, deduced from the protons in coincidence with the Si recoils. 
A $Q$-value resolution of approximately 150~keV FWHM was achieved. Four known, low-lying states of $^{33}$Si~\cite{Jongile,Wang} are clearly identified in the spectrum, corresponding to the ground ($3/2^+$), 1.01-MeV ($1/2^+$), 1.435-MeV ($7/2^-$), and 1.981-MeV ($3/2^-$) states. They are associated with the transfer of a neutron into the $0d_{3/2}$, $1s_{1/2}$, $0f_{7/2}$ and $1p_{3/2}$ orbitals, respectively. Two new states are observed at 3.19(2) and 3.58(2)~MeV, below the neutron-separation energy ($S_n=4.508$~MeV). There are also two weakly populated resonances observed at around 4.52 and 5.43 MeV. The differential cross sections measured for the observed states are shown in Fig.~\ref{fig3}.  Each PSD on the array is divided into one or two angular bins depending on statistics. %
Relative cross sections used in the following analysis have a systematic uncertainty of around $5\%$, which is dominant by the angular range covered by the silicon array, and the cut on the $\Delta E$-$E$ recoil detectors. The absolute cross sections were normalized to the elastic scattering events in the recoil detectors.  The uncertainties of the absolute cross sections are estimated to be around 30-50$\%$, which is dominated by the estimated uncertainty in the angular coverage of the recoil detectors. At these forward center-of-mass angles, changes in a few tenths of a degree can modify the calculated cross section by as much as 50$\%$.

Distorted-wave Born approximation (DWBA) calculations were performed with the code \textsc{ptolemy}~\cite{PTOLEMY}. Optical-model parameters (OMPs) of Refs.~\cite{Koning,An} were used. For the four lowest-lying states, the agreement between the experimental angular distributions and the DWBA calculation confirms previous $\ell$ assignments. The newly observed 3.58(2)-MeV state has an $\ell=1$ shape, which may be associated with a neutron transfer into the $1p_{1/2}$ or $1p_{3/2}$ orbitals as discussed below. 
A tentative assignment of $\ell=3$, and an assumption of a $0f_{7/2}$ orbital, is made for the smaller peak at 3.19(2) MeV. For the unbound states, a binding energy of $200$ keV for the transferred neutron was assumed in a “quasi-bound” approach. 
The 4.52(4)- and 5.43(4)-MeV resonances have fitted widths $\Gamma=220(80)$~keV and $\Gamma<90$~keV, and were tentatively assigned $\ell=1,2$ and $\ell=3$, respectively.

\begin{table}
\caption{\label{tab:expt-S} Excitation energies $E_x$, transferred orbital angular momentum $\ell$, spin-parities $j^{\pi}$,  shell-model orbital $n \ell j$ and normalized spectroscopic factors $S$ for the low-lying states in $^{33}$Si  observed in the  $^{32}$Si\,$(d,p)^{33}$Si reaction. }
\newcommand\T{\rule{0pt}{3ex}}
\newcommand \B{\rule[-1.2ex]{0pt}{0pt}}
\begin{tabular}{lcccc}
\hline
\hline
\textrm{$E_x$\,(MeV)}&
\textrm{$\ell$}&
\textrm{$j^{\pi}$}&
\textrm{$n \ell j$}&
\textrm{$S$}\\
\hline
\T g.s.   & 2 & $3/2^+$ &  $0d_{3/2}$&0.37(4)\\
1.01 & 0 & $1/2^+$ & $1s_{1/2}$ &0.25(5)\\
1.435 & 3 & $7/2^-$ & $0f_{7/2}$ &0.89(5)\\      
1.981 & 1 & $3/2^-$ & $1p_{3/2}$ &$0.92(6)$\\    
3.19(2) & (3) & ($7/2^-$) & ($0f_{7/2}$) &0.07(2)\\   
3.58(2) & 1 & $1/2^-$ & $1p_{1/2}$ &0.91(7)\\
\multirow{2}{*}{4.52(4)}&(1)&$(3/2^-/1/2^-)$&$(1p_{1/2,3/2})$&0.08(2)\\
            & (2) & $(3/2^+/5/2^+)$ & $(0d_{5/2,3/2})$ & 0.10(3) \\
5.43(4) & (3) & $(7/2^-/5/2^-)$ & $(0f_{7/2,5/2})$ &0.10(3)\\
\hline
\end{tabular}
\end{table}

Since the more bound $0d_{5/2}$ orbit is almost full, the $1s_{1/2}$ and $0d_{3/2}$ orbitals have two shared vacancies  in $^{32}$Si, with $N=20$ being the closed shell. The relative spectroscopic factors were thus normalized so that their summed strength $\sum(2j+1)C^2S$ is $2.0$. The same normalization factor was then also applied to the $\ell=1$ and 3 states yielding the normalized spectroscopic factors listed in Table~\ref{tab:expt-S}. The uncertainty of the relative spectroscopic factors was dominated by the variation of the OMPs, which is less than 10$\%$~\cite{Kay2013}. The relative spectroscopic factors of the 1.435-MeV ($7/2^-$), 1.981-MeV ($3/2^-$) and the newly observed 3.58-MeV state are close to 1.0, which is commensurate with the expected full single-particle strength of the nominally empty neutron $1p$ and $0f_{7/2}$ orbitals.
This sum-rule analysis strongly supports a $1/2^-$ assignment to the 3.58(2)-MeV state since it almost exhausts the full $1p_{1/2}$-orbital single-particle strength. This analysis suggests the dominant fraction of $1p_{1/2,3/2}$ and $0f_{7/2}$ orbital single-particle strengths are observed below 6 MeV, similar to $^{35}$Si.

\begin{figure}
\includegraphics[width=1.0\columnwidth]{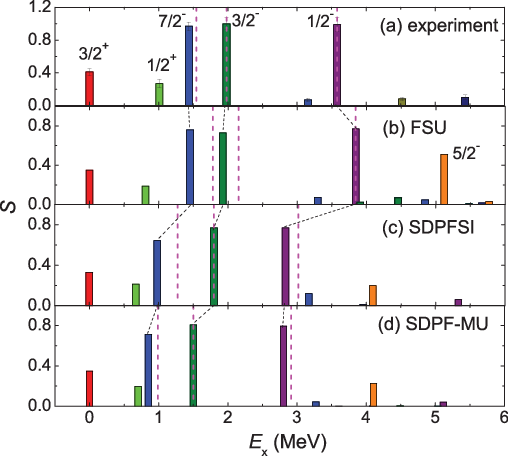}
\caption{\label{fig4}Excitation energies and corresponding spectroscopic factors of the low-lying states in $^{33}$Si measured in the $^{32}$Si($d,p$)$^{33}$Si reaction compared to shell-model calculations using the FSU, SDPF-SI and SDPF-MU interactions. 
The dashed pink lines are the centroids of the $7/2^-$, $3/2^-$, and $1/2^-$ states.} 
\end{figure}
\begin{figure*}[hbt]
\centering
\includegraphics[width=2.0\columnwidth]{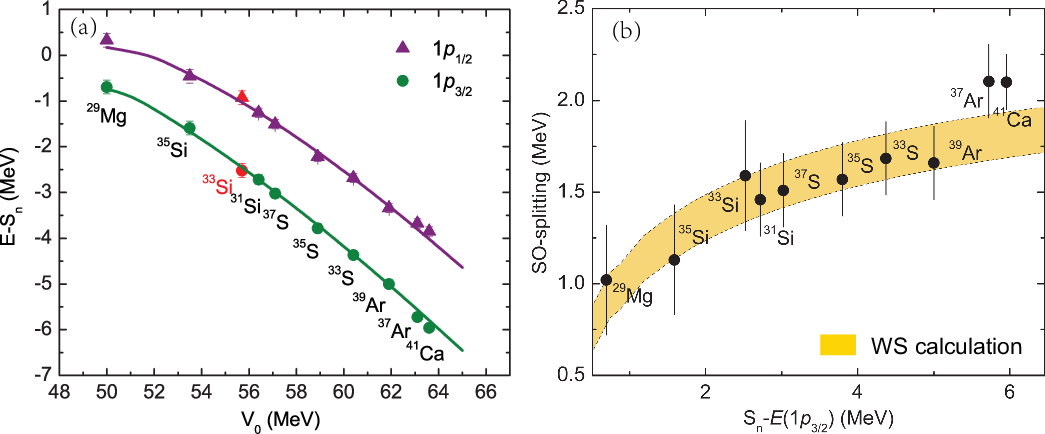}\\
\caption{(a) Experimental binding energies of the neutron $1p_{3/2}$ (green circles) and $1p_{1/2}$ (purple triangles) orbitals as a function of the fitted depth of the Woods-Saxon potential and compared to the calculated values (solid curves) with fixed geometry. The red symbols highlight the $^{33}$Si data. (b) SO-splitting of the neutron $1p$-orbitals  as a function of the corresponding neutron $1p_{3/2}$-orbital binding energies. The shaded band shows the result of the Woods-Saxon calculation with the associated uncertainties. } \label{spe}
\end{figure*}

\section{Discussion}
\label{Discussion}

The excitation energies, spectroscopic factors, and SPE centroids of the $^{33}$Si low-lying states are compared with shell-model calculations using FSU~\cite{Lubna}, SDPF-SI~\cite{SDPF-SI} and SDPF-MU~\cite{SDPF-MU} interactions in Fig.~\ref{fig4}. In these calculations, a model space allowing for one particle to move across the $N=Z=20$ shell gap ($0$--$1\hbar\omega$) was used, without the mixing between 0p-0h and 2p-2h or 1p-1h and 3p-3h configurations. The SDPF-SI and SDPF-MU interactions underestimate the excitation energies of the low-lying states, while the FSU interaction reproduces them reasonably well. Since the experimental data are reproduced without significant configuration mixing, the $N=20$ shell gap is observed to persist in $^{32,33}$Si as expected from previous measurements~\cite{Fortune,Ibbotson}. The SPEs of the neutron $0f_{7/2}$ and $1p_{3/2, 1/2}$ orbitals are determined from the spectroscopic-factor-weighted average energy of states with a given $j$~\cite{Baranger}.  From the calculations, any fragments of single-neutron strengths outside of these lowest-lying states shift their centroid energies at most by $\sim$250 keV. The shell model calculations predict that about 90$\%$ of the ideal sum-rule value strength is obtained below 6 MeV for each orbital. The lowest $7/2^-$, $3/2^-$ and $1/2^-$ states should account for $90\%$ of the predicted single-particle strength.

\begin{figure}
\includegraphics[width=1.0\columnwidth]{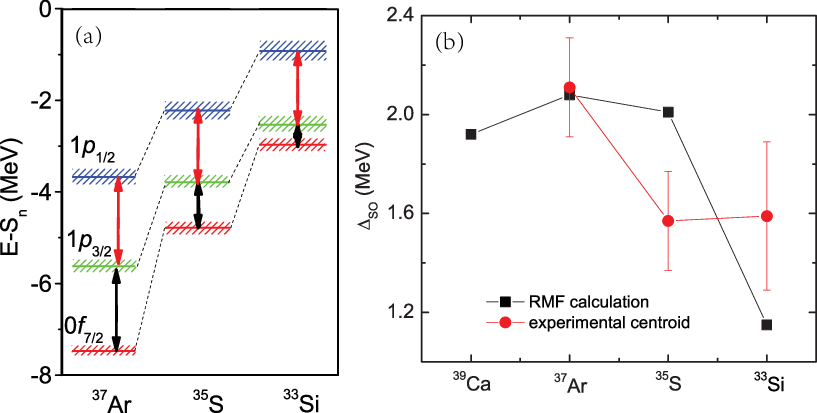}\\
\caption{\label{fig5} (a) Binding energies of the neutron $1p_{1/2}$  (blue), $1p_{3/2}$ (green) and $0f_{7/2}$ (red) orbitals in the $N=19$  isotones as determined from the data. The slashed areas indicate their uncertainties.   (b) The SO splitting $\rm \Delta_{SO}$ of the $1p$-orbitals of $N=19$ isotones predicted by the RMF theory in comparison with the experimental values determined from the centroid. 
} 
\end{figure}

The experimental binding energy of the $0f_{7/2}$, $1p_{3/2}$ and $1p_{1/2}$ orbital was determined to be -2.95(15), -2.53(15) and -0.93(15) MeV, respectively, according to the method in Ref.~\cite{Baranger}. The SO splitting is consequently 1.60(30)MeV. 
Experimentally, no significant fragmentation of the $\ell=1$ strength was observed, which is supported by the shell-model calculation above. Therefore, the binding energies of the $1p_{3/2}$ and $1p_{1/2}$ orbitals were determined by the lowest $3/2^-$ and $1/2^-$ states, respectively. The possible $\ell = 1$ resonance at 4.52 MeV would shift the SPEs by at most 150~keV, which has been incorporated in the uncertainties. The $0f_{7/2}$ single-particle energy is determined by taking the weighted average of the 1.435- and 3.19-MeV states. The SPEs of these orbitals in the neighboring $N=19$ isotones $^{35}$S~\cite{Piskor,Mermaz} and $^{37}$Ar~\cite{Sen,VanDer} were also determined from existing data [Fig.~\ref{fig5}(a)]. The $1p_{3/2}$ and $0f_{7/2}$ SPEs of $^{37}$Ar have been shifted downward by around 100 and 250~keV, respectively, when the neutron-removal strength was considered~\cite{Baranger}. The $pf$-shell orbitals of $^{35}$S have been shifted by less than 50~keV. For $^{33}$Si, the removal strength impact is expected to be no greater than in $^{35}$S or $^{37}$Ar. 

In Fig.~\ref{spe}(b), the neutron $1p$-orbital SO-splitting ($\rm \Delta_{SO}$) of $N=17$, 19, and 21 isotones reconstructed from the current measurement and literature data are plotted as a function of the corresponding neutron $1p_{3/2}$ SPE. The $1p$ SPEs of the $N=17$ and 21 isotones are taken from Refs.~\cite{Kay,MacGregor,nndc}. The uncertainties vary case by case, but most are within 100-300 keV. There is a strikingly clear, smooth trend in $\rm \Delta_{SO}$ as a function of binding energy. The data for both $^{35}$Si and $^{33}$Si lie along this smooth trend, together with their sulfur counterparts $^{37}$S and $^{35}$S, so there is no evidence of a sudden reduction in the SO-splitting from $Z=16$ to $Z=14$. The smooth dependence on the binding energies is an indication that the finite-binding effect may play a significant role. 

The evolution of the $1p$ SPEs can be described by a simple Woods-Saxon potential, including data in the region approaching zero neutron binding energy.  Fig.~\ref{spe}(a) shows the binding energy of the $1p_{1/2}$ and $1p_{3/2}$ orbitals, as a function of the fitted depth of a Woods-Saxon potential using the potential parameters $r_0=1.2$ fm, $a_0=0.7$ fm, $r_{so}=1.3$ fm, $a_{so}=0.65$ fm and $V_{so}=6$ MeV. The depth of the potential was chosen to reproduce the binding energies of these two orbitals using the $\chi^2$ minimization method. The SO strength is not varied in the calculation. It is immediately apparent that the SO splitting and SPEs of the neutron $1p$ orbitals are reproduced by the calculation without a need for modification of the SO strength. A range of sensible WS parameters were investigated but with the same general conclusion. Agreement between the data and calculation indicates that the smooth evolution of the neutron SO splitting follows the previously noted {\it lingering} effect of the low-$\ell$ orbitals~\cite{Kay}, which is a direct effect of the extended nature of their wave functions.

Development of a proton bubble structure in $^{34}$Si requires two attributes: a very small proton occupancy in the $1s_{1/2}$ orbital and little-to-no coherent correlations between the nucleons. The $1s_{1/2}$ proton occupancy of $^{34}$Si has been determined to be 0.07(3) (compared to its isotone $^{36}$S $\approx$ 1.8)~\cite{Mutschler}, while in $^{34}$S it is 1.4-2.0~\cite{nndc}.  The deformation magnitude of the $^{32}$Si first excited 2$^+$ level is unexpectedly small, which is well below theoretical predictions~\cite{Ibbotson,32SiCoulomb}. The latter information is a strong indication that the protons form a good $Z$=14 core, as supported by shell-model calculations. 

In the RMF calculation with the DD-ME2 interaction~\cite{Lalazissis}, $^{32}$Si was predicted to exhibit a central density depletion, similar to $^{34}$Si, due to low $1s_{1/2}$ proton occupancy. This calculation predicts a sudden reduction of the neutron $1p$ SO splitting ($\rm \Delta_{SO}$(S)-$\rm \Delta_{SO}$(Si)$\approx$0.9 MeV) in $^{33}$Si compared to $^{35}$S, similar to the $N$=21 isotones. However, from the present measurement, there is little reduction of SO splitting in $^{33}$Si compared to $^{35}$S ($\rm \Delta_{SO}$(S)-$\rm \Delta_{SO}$(Si) $\approx$ -0.1\,MeV), which is in contradiction to the RMF calculation (see Fig.~\ref{fig5}(b)). The mismatch of this calculation might be attributed to the fact that the proton-neutron quadrupole correlations are not taken into account in the RMF calculation. 
Therefore, from the consistency of the empirical $\rm \Delta_{SO}$ trend and contradiction with the RMF calculation, the existence of a sudden reduction of SO splitting associated with a proton bubble is not supported. 
It is noted that the $\rm \Delta_{SO}$ of $^{29}$Mg is the smallest among these nuclei, which cannot be explained by the presence of a proton bubble.

The SO coupling is a surface term by definition~\cite{Bohr}. By approximating the SO potential to a $\delta$ function at the nuclear surface, a simple evaluation of the SO splitting was established~\cite{Bohr, Orlandi}, $\rm \Delta_{SO} \propto V_{so}(\boldsymbol{\ell} \cdot \boldsymbol{s}) r_{0}^{2} R \Psi^{2}(R)$, where $\rm V_{so}$ is the SO potential strength, $\rm \Psi(R)$ is the radial wave function and $\rm R$ is the nuclear radius.   
Due to the finite binding effect, the wave functions of the neutron $1p$ orbitals have smaller surface radial amplitude when becoming weakly confined. Using the calculation with a WS potential, it is found that the $\rm R \Psi^{2}(R)$ term reduces gradually, lower than 60\% of its original value when the binding energy decreases from 2.9 to 0.1 MeV. Therefore, the apparent SO-splitting reduction can be accounted for by the evolution of the neutron $1p$ wave functions at the surface. 

The dramatic narrowing of the $N=28$ shell gap can also be inferred from Fig.~5(a), seen from the change in separation of the $0f_{7/2}$-orbital binding energy below $N=28$ and that of the $1p_{3/2}$ orbital above it. The relative energy reduction of the $1p_{3/2}$ orbital is in part due to the differing behaviors of the $1p_{3/2}$ and $0f_{7/2}$ orbitals as they become less bound; the lingering effect is more pronounced for the low-$\ell$ orbitals. 

\section{Summary}
\label{Summary}

In conclusion, the SPEs of the neutron $1p_{1/2}$,  $1p_{3/2}$ and $0f_{7/2}$ orbitals have been discussed for the neutron-rich $N=19$ isotones, including new data on $^{33}$Si. Combined with the neutron $1p$-shell SPEs in the $N=17$ and $N=21$ isotones,  a smooth reduction in the SO splitting is found when nuclei become less bound; this feature can be reproduced by a calculation with a WS potential without any modifications of the SO strength. These phenomena agree with an interpretation of the SO-splitting evolution resulting from the geometric effect of the nuclear potential. 
Further insight may be gained from a systematic mapping of the SO splitting across the region via one-nucleon transfer-reaction experiments, which is an exciting prospect with modern-day facilities. 

\section{Acknowledgements}
\label{Acknowledgements}

The authors would like to acknowledge the operation staff at ReA6 (NSCL) for providing the beam. This material is based upon work supported by National Superconducting Cyclotron Laboratory, which has been a major facility fully funded by the National Science Foundation under award PHY-1565546; the U.S.\ Department of Energy, Office of Science, Office of Nuclear Physics, under Contract Number DE-AC02-06CH11357 (Argonne), DE-SC0020451 (FRIB) and under Award Number DE-SC0014552 (UConn); NSF grant PHY-2011398; the Spanish Ministerio de Economía y Competitividad through the Programmes “Ramón y Cajal” with the grant number RYC2019-028438-I; the U. K.\ Science and Technology Facilities Council (Grant No. ST/P004423/1, ST/R004056/1 and ST/T004797/1); the International Technology Center Pacific (ITC-PAC) under Contract No. FA520919PA138 and Australian Research Council Grant No. DP210101201. SOLARIS is funded by the DOE Office of Science under the FRIB Cooperative Agreement DE-SC0000661. FSU shell-model calculations were performed by using the computational facility of Florida State University, which is supported by the grant number DE-SC0009883 (FSU). We gratefully acknowledge the use of the Bebop cluster in the Laboratory Computing Resource Center at Argonne National Laboratory. Data associated with this experiment can be obtained by reasonable request to the author.



\end{document}